\documentclass[aps,prb,twocolumn]{revtex4}
\usepackage{graphicx}
\usepackage{dcolumn}
\usepackage{bm}
\usepackage{amssymb}
\usepackage{times}
\usepackage{hyperref}
\usepackage{amsmath}
\usepackage{units}
\usepackage{nicefrac}
\begin{document}

\title{Metamaterial Coatings for Broadband Asymmetric Mirrors}
\author{A. Chen, K. Hasegawa and M. Deutsch}
\address{Department of Physics, University of Oregon, Eugene, OR 97403}
\author{V.A. Podolskiy}
\address{Department of Physics, Oregon State University, Corvallis, OR 97331}
\date{\today}

\begin{abstract}
We report on design and fabrication of nanostructured metamaterial metal-dielectric thin film coatings with high reflectance asymmetries. Applying basic dispersion engineering principles to model a broadband and large reflectance asymmetry, we obtain a model dielectric function for the metamaterial film, closely resembling the effective permittivity of disordered metal-dielectric nano-composites. Coatings realized using disordered nanocrystalline silver films deposited on glass substrates confirm the theoretical predictions, exhibiting symmetric transmittance, accompanied by unique large broadband reflectance asymmetries.
\end{abstract}

\maketitle

 \noindent Optical metamaterials -- artificial composites with
engineered electromagnetic (EM) response -- hold the potential for
greatly impacting modern photonics, with promising applications as
novel guiding, imaging and dispersive optical
components\cite{scalora,superlens,hyperlens,waveguide}. The two main
classes of metamaterials comprise disordered metallodielectric
nanocomposites~\cite{Mulvaney,Ding} and materials with long range
ordering such as photonic crystals~\cite{Baumberg}. In this Letter
we demonstrate an application of disordered metamaterials which
harnesses their dispersive properties. In particular, we implement
disordered metal--dielectric films to achieve highly asymmetric
broadband optical reflectors.

An asymmetric mirror is a planar layered optical device exhibiting asymmetry in reflectance of light incident from either side, while its transmittance is symmetric~\cite{Kard}. In this system the energy-balance relations are $T+R_{1,2}+A_{1,2}=1$, where $T$, $R$ and $A$ are the mirror transmittance, reflectance and losses (in form of absorption as well as scattering,) respectively, and the subscripts 1 and 2 specify the direction of light-incidence on the mirror. Asymmetric mirrors have recently found use in specialized Fabry-Perot interferometers~\cite{Troitskii1}. To obtain such a mirror, two conditions must hold: (i) The structure must lack inversion symmetry, and (ii) the mirror should impart a non-unitary energy-transformation to the beam. The latter may be achieved through out-of-beam scattering, or by assuring that the dielectric function of at least one of the films is complex, i.e. exhibiting either absorptive losses \emph{or} gain. One of the simplest structures for an asymmetric mirror is a thin metal film on a dielectric slab, embedded in a uniform dielectric.

Asymmetric mirrors have been realized using smooth metal films on planar dielectric substrates~\cite{Goldina,Troitskii2}, or metal gratings~\cite{Zeng}. The optical response of these mirrors, such as their reflectance asymmetry and associated bandwidth are typically constrained to a narrow range, due to a limited choice of materials. We show that these characteristics may be dramatically enhanced in \emph{metamaterials}.

Solving Maxwell's equations for an EM field of frequency $\omega$ impinging on a film of thickness $d$ and permittivity $\epsilon_f$, deposited on a semi-infinite substrate embedded in vacuum, the reflectance asymmetry is

\begin{equation}
\Delta R\equiv R_1-R_2=\frac{\left|AB+C\right|^{2}-\left|AC+B\right|^{2}}{\left|1+ABC\right|^{2}}
\label{eq1}
\end{equation}

\noindent In this notation $A\equiv e^{2ik_f d}$, with $k_f=\sqrt{\epsilon_f}\omega/c$ and $c$ the speed of light in vacuum. We limit our discussion to absorptive films, hence $\epsilon_f\equiv \epsilon_f'+i|\epsilon_f''|$. The amplitude reflection coefficients  $B\equiv r_{12}$ and $C\equiv r_{23}$ denote reflections from the vacuum--metal and metal--dielectric interfaces, respectively, and are given by standard textbook expressions. The numerator of Eq.~(\ref{eq1}) may be rewritten as $\left(\left|A\right|^{2}-1\right)\left(\left|B\right|^{2}-\left|C\right|^{2}\right)+
2Re\left[A\left(BC^{*}-B^{*}C\right)\right]$. From this we see that $\Delta R=0$ either in systems with inversion symmetry where $B=-~C$, or for lossless materials, where either $|A|=1$ and $B$ and $C$ are \emph{real} (as in lossless dielectrics), or $|B|=|C|=1$ and $A\left(BC^{*}-B^{*}C\right)$ are \emph{pure imaginary} such as in lossless metals.

\begin{figure}[b]
\centering
\includegraphics[width=8.5 cm ]{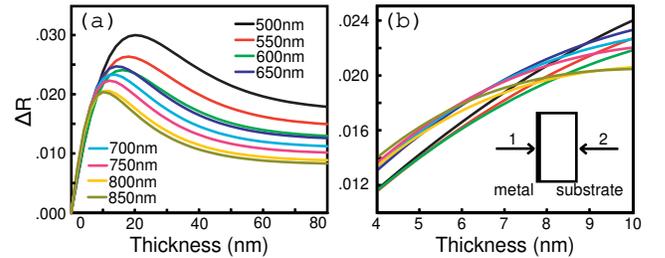}
\caption[The dependence of $\Delta$R on metal film
thickness.]{\label{fig1}(a) The dependence of $\Delta$R on metal
film thickness. (b) Magnified view of the near-crossover in (a).
Inset: Schematic of the asymmetric mirror, with numbers denoting the
direction of light incidence.}
\end{figure}

It is important to understand the dependence of the reflectance asymmetry on the optical constants of the metal film, and in particular its dependence on optical losses. Such knowledge is instrumental in designing asymmetric mirrors with controllable spectral response. Since in general the permittivity of the metal is a function of frequency, a simple closed-form expression for $\Delta R$ is not always available. In Fig.~\ref{fig1}(a) we plot the reflectance asymmetry of a thin silver film on a glass slab embedded in vacuum, calculated using experimentally tabulated values for silver~\cite{Palik}. Close examination of $\Delta R$ in Fig.~\ref{fig1}(b) reveals frequency dependence, resulting from the aforementioned material dispersion as well as finite film-thickness effects, while the value of $\Delta R$ in the minimum dispersion range is only $\sim 2\%$. This behavior is typical to most metals which are reflective in the visible and near-infrared (NIR). Certain applications (e.g. in solar cells) may benefit from increasing the value of $\left|\Delta R\right|$, with simultaneous minimization of its dispersion. This requires careful design of the film's structure and composition, yielding a dispersion-engineered metamaterial. Below we present an example of such a design process.

\begin{figure}[t]
\centering
\includegraphics[width=8.5 cm ]{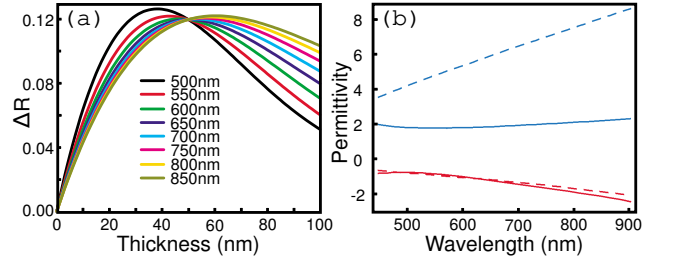}
\caption[Engineered reflectance asymmetry and the
permittivity]{\label{fig2} (a) Engineered reflectance asymmetry,
showing an exact crossover at $d=50$nm where $\Delta R$ is
non-dispersive. (b) Solid lines: real (red) and imaginary (blue)
components of the permittivity obtained from inverting $\Delta R$ in
(a). Dashed lines: best-fit of $\epsilon_f'$ and $\epsilon_f''$
obtained using Bruggemann EMT.}
\end{figure}

We start by imposing two constraints: (i) The asymmetry should be large ($\sim10\%$) and (ii) the asymmetry should posses a \emph{broadband} characteristic, i.e $\partial\left(\Delta R \right)/\partial\lambda =0$ at a given film thickness. Note that the thickness of thin metamaterial films is typically constrained by the fabrication, and therefore is not always a good variable in the design process. To illustrate our approach, we select the functional form of $\Delta R$ to resemble that in Fig.~\ref{fig1}(a). The broadband condition stated above implies the existence of a design parameter -- the film thickness, $d$ -- at a particular value of which $R$ is constant over a wide range of wavelengths. This (exact) crossover point in $R$ is chosen here to occur at $d=50nm$.  Fig.~\ref{fig2}(a) shows the desired asymmetry of the thin metamaterial film, plotted against film thickness over the visible and NIR range. The permittivity $\epsilon_f(\lambda)$ may now be extracted from the expression for $\Delta R$ by simple inversion. In general, for a given shape of $\Delta R$, there exists a wide range of values of $d$ ($0<d\simeq50$nm) and $\Delta R$ (up to $\sim15\%$) for which the resulting $\epsilon_f(\lambda)$ exhibits the general form expected for a material satisfying the causality relations.

The resulting components of the complex dielectric response are shown in Fig.~\ref{fig2}(b). We see that $\epsilon_f'$ is negative over the entire visible range, implying a metallic response. For large enough values of $\Delta R$ ($\sim15\%$) $\epsilon_f'>0$ at short wavelengths, crossing over to negative values only in the mid-visible range. Comparing the desired properties of $\epsilon_f$ to permittivities of noble metals, we note that (i) $\epsilon_f''$ is significantly greater than the known values for silver, and (ii) the effective plasma wavelength of $\epsilon_f$ is red-shifted with respect to that of silver.

While the desired $\epsilon_f$ differs from permittivities of known materials, a dielectric function similar to the one in Fig.~\ref{fig2}(b) may be achieved in a metamaterial, often described by effective medium theory (EMT). According to EMT, composite materials with spatial inhomogeneities of typical size much smaller than the relevant wavelength may be regarded as homogeneous on average. To demonstrate our design we apply Bruggemann EMT\cite{Bruggeman} to model a nano-composite film with metal filling fraction $p$. The effective dielectric function $\epsilon_{\hbox{\scriptsize eff}}$ is given by

\begin{equation}
p\frac{\epsilon_m-\epsilon_{\hbox{\scriptsize
eff}}}{g\epsilon_m+(1-g)\epsilon_{\hbox{\scriptsize eff}}}
+(1-p)\frac{\epsilon_d-\epsilon_{\hbox{\scriptsize
eff}}}{g\epsilon_d+(1-g)\epsilon_{\hbox{\scriptsize eff}}}=0
\label{eq2}
\end{equation}

\noindent where $\epsilon_m$ and $\epsilon_d$ are known dielectric functions of the metal and the dielectric, respectively, and $g=0.68$ is a constant describing the microscopic morphology of the film's constituents, and is also known as the depolarization factor\cite{Cohen}. Here we have introduced \emph{a second design parameter} in the form of $p$~\cite{footnoteMicro}. Ultimately, this approach allows the design of composite materials with prescribed dispersion under a given set of fabrication constraints.

The dashed lines in Fig.~\ref{fig2}(b) show the results of EMT modelling of a silver nano-composite embedded in vacuum. We find that a value of $p=0.71$ yields excellent agreement with the desired $\epsilon_f'$. The discrepancy in the modelling of $\epsilon_f''$ is due to the microscopic loss mechanisms specific to the Bruggemann model, which cannot exactly reproduce the desired response.

\begin{figure}[t]
\centering
\includegraphics[width=3.1in]{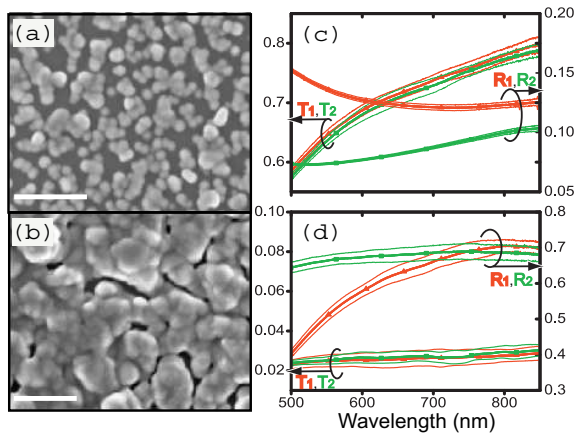}
\caption[Sample characterization and the corresponding reflectance
and transmittance]{\label{fig3} (color online) Left: SEM images of
films with different metal ratios: (a) $p=0.52$ (b) $p=0.93$. Scale
bars are 400nm. Right (c), (d): Measured reflectances and
transmittances of the films at left. Heavy lines are measured
values; thin lines denote the range of error, obtained from signal
averaging.}
\end{figure}

Semi-continuous silver films with varying filling fractions were deposited on microscope glass slides using a modified Tollen's reaction~\cite{Emmons2006}. The typical morphologies of the films achieved by this method can be seen in the scanning electron micrographs (SEMs) in Fig.~\ref{fig3}. The degree of surface coverage was controlled by monitoring the chemical deposition time, with reactions lasting from $1-6$ hours. The metal filling fractions ranged from $p\approx 0.1-0.9$ and were determined from high resolution SEM images, using a previously developed image analysis method\cite{Rohde}. Ensuing deposition all samples were stored under inert conditions, to minimize silver oxidation.

Optical reflectance and transmittance spectra were collected using an inverted optical microscope whose output port was directly imaged on the entrance slit of a 320mm focal length spectrometer. Reflected and transmitted light signals from a tungsten-halogen white light illuminator impinging at normal incidence on each side of the sample were collected with a 10X objective (0.25 N.A.)\cite{footnoteNA} and imaged onto a liquid-nitrogen-cooled charge-coupled device. The reflected signals were carefully normalized using a high-reflectance mirror (Newport Broadband SuperMirror, $R\geq$99.9\%.) To eliminate spurious effects from local inhomogeneities, signals collected from $\sim1000$$\mu$m across the film were averaged. Various degrees of reflectance asymmetry were observed for films with different silver filling fractions. Nevertheless, the transmittance \emph{always} remained symmetric, even for rough films with high surface coverage, as shown in Fig.~\ref{fig3}(c)-(d). The latter indicates that the transmission symmetry is not broken by the disorder-mediated (i.e. \emph{diffuse}) scattering from the rough silver interfaces.

\begin{figure}[t]
\centering
\includegraphics[width=3.0in]{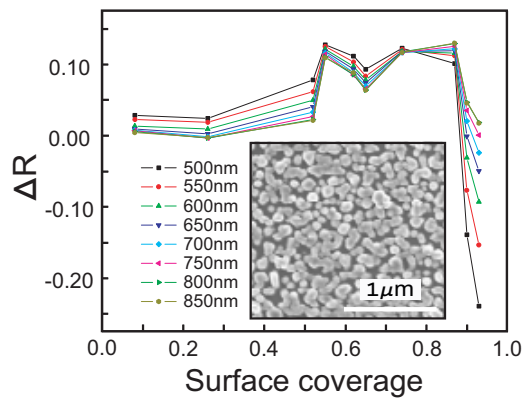}
\caption[Reflectance asymmetry from different samples]{\label{fig4}
(color online) Reflectance asymmetry measured as function of $p$.
Inset: SEM micrograph of $p=0.74$ film.}
\end{figure}

It is now possible to address the effect of metal filling fraction on the reflectance asymmetry. In Fig.~\ref{fig4} we plot $\Delta R$ for 10 samples with increasing values of $p$. As can be seen, both the magnitude and sign of $\Delta R$ depend strongly on surface coverage. Comparing this result to the predicted model in Fig.~\ref{fig2}(a) yields good agreement for the general shape and trend of $\Delta R$, as well as for its target value of $\Delta R\sim10\%$. We note that instead of plotting $\Delta R$ as function of film thickness we plot it against $p$, as the latter can be measured with much higher accuracy than the thickness of these rough films. The most noticeable feature is the theoretically predicted crossover point near $p=0.74$, where the dispersion in $\Delta R$ is minimal.

Comparing the data in Fig.~\ref{fig4} to Fig.~\ref{fig2}(a) we find that the functional form of $\Delta R$ in experiment differs from theory, especially near $p=0.6$, where $\Delta R$ is non-monotonic. This is not surprising, since EMTs such as the Bruggemann approach used to model $\epsilon_f$ in Fig.~\ref{fig2}(b), while resembling the optical response of metamaterials, do not provide a complete description of scattering at rough interfaces. In particular, scattering boundary conditions and film thickness are usually poorly known for discontinuous metal films, whose optical response is dominated by surface scattering and enhanced absorption. Indeed, we have computed $\epsilon_f$ of our films using the measured filling fractions and Eq.~(\ref{eq2}), but were not able to reliably reproduce the asymmetry for most values of $p$.

It is now possible to utilize our approach to design metamaterial coatings with extended functionalities. Loss-compensated asymmetric mirrors may be realized by incorporating a gain medium in the embedding matrix\cite{Noginov}. Alternately, electro-active or semi-conducting matrices may enable implementation of these mirrors in photonic devices, solar cells and full-color displays.

In summary, we realized strongly asymmetric mirrors using disordered nanocrystalline silver films deposited on glass substrates. Basic dispersion engineering principles were applied to model a broadband and large reflectance asymmetry, which was then inverted to yield the effective permittivity. An effective-medium approach was then
implemented to approximate the required optical response function in a metal-dielectric metamaterial, closely mimicking that of disordered silver-dielectric composites. The dependence of the optical asymmetry on metal filling fraction in the coatings was measured, demonstrating the predicted broadband characteristic.

This work was supported by National Science Foundation grant DMR-02-39273.


\begin{thebibliography}{20}
\bibitem{footnoteMicro}While $p$ is the only EMT parameter for disordered films, the response of other classes of metamaterials may also depend on additional parameters such as shape (i.e. $g$), size, order and permittivities of the constituents. The parameters describing these characteristics will impose additional constraints on the design process.
\bibitem{footnoteNA}Calculations show $\Delta R$ depends weakly on angle near normal incidence, thus the experimental conditions are well-approximated by the plane-wave modeling applied here, and no further angular averaging was necessary.
\end{thebibliography}
\end{document}